\documentstyle[prl,aps,twocolumn,psfig]{revtex}
\begin{document}

\title{Angular Correlations in Breakup of Three--Body Halo Nuclei }
\author{E. Garrido}
\address{Instituto de Estructura de la Materia, CSIC, Serrano 123, E-28006
Madrid, Spain}
\author{D.V.~Fedorov and A.S.~Jensen}
\address{Institute of Physics and Astronomy,
Aarhus University, DK-8000 Aarhus C, Denmark}
\date{\today}

\maketitle

\begin{abstract}
We use the three-body model and the sudden approximation to compute
angular correlations in high-energy fragmentation reactions of
two-neutron halos on light targets. The contribution from one-neutron
absorption by far dominates over that of neutron scattering.  We use
$^6$He (n+n+$\alpha$) and $^{11}$Li (n+n+$^{9}$Li) as examples and
study the dependence of the predictions of this model on different
physical assumptions and parameters.~\\
\indent {PACS number(s): 25.60.-t, 25.60.Gc, 21.45.+v}~\\
\end{abstract}

{\it Introduction.} A new class of nuclear states, called halos, was
discovered about ten years ago \cite{han95}.  Halos are spatially
extended states with small one or two-nucleon separation energies and
low orbital angular momenta. The systems can be understood essentially
as few-body systems where we can separate the coordinates into tightly
bound intrinsic (core) and loosely bound external (halo) degrees of
freedom. The characteristic properties, unusual in the nuclear
context, are related to the few halo degrees of freedom while the core
can be assumed to be inert. Prominent examples are the two-neutron
halos consisting of two neutrons and a core. The properties of halo
systems have been remarkably well described by three-body models
\cite{han95,zhu93,gar96,gar97,gar98}.  The lesson is clearly that properties
related to the halo degrees of freedom (the core remains intact) can
be deduced from three-body models.

However, the recently measured neutron angular correlations in
fragmentation reactions of Borromean two-neutron halos are apparently
sensitive to both the reaction mechanism and the halo structure
\cite{chu97}.  A satisfactory description is not available. Other
previously measured neutron momentum distributions in high-energy
breakup reactions have been reproduced in a three-body model where the
sudden approximation and the final state interactions are decisive
ingredients \cite{gar96,gar97}. Recently this model was extended to
include both absorption and diffraction of the halo particles on the
target and in addition also the absolute two-neutron removal cross
sections were successfully calculated \cite{gar98}.  The purpose of
this report is to analyze the recently measured neutron angular
correlations within this new model. We shall study the dependence on
various physical parameters, compare to available data, see how far
the model can go, possibly suggest new reaction mechanisms and predict
yet unobserved neutron correlations.

{\it Model and method.} The spatially extended three-body halo
collides with a relatively small target at high energy. Then the
probability that more than one of the constituents interacts strongly
with the target is very small. The differential cross section
$d\sigma$ is then to a good approximation a sum of three terms
$d\sigma ^{(i)}$ each describing the independent contribution to the
process from the interaction between the target and the halo particle
$i$.  This is the assumptions used in the classical formulation for a
weakly bound projectile \cite{ban67}. We neglect the binding energy of
the initial three-body bound state compared to the high energy of the
beam.  The reaction is then described as three particles independently
interacting with the target as if each particle was free. 

The process is described as removal of one particle (participant)
while the other two particles (spectators) both survive undisturbed.
The participant is either absorbed or elastically scattered by the
target.  The final state then consists of two independent subsystems,
i.e. the two spectators and the target plus participant. The final
state interaction between spectators is necessarily the same as in the
initial three-body bound state which is described in agreement with
available experimental information \cite{gar96,gar97}.  Since we only
need to account for details of the scattered particle we employ a
phenomenological optical potential where all other processes are
included as absorption from the elastic channel \cite{udi95}.

\begin{figure}
\psfig{file=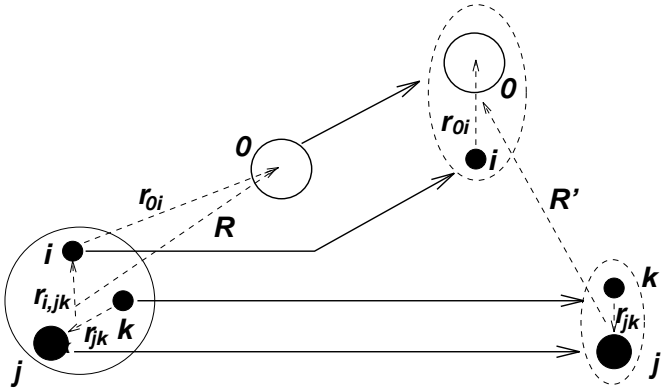,angle=-90,width=87mm}
\vspace{3mm}
\caption{Sketch of the reaction and the coordinates used. The
target is labelled by 0 and $\{i,j,k\}$ label the particles within the
three-body projectile.}
\label{fig1}
\end{figure}

The coordinates, ${\bf r}_{jk}, {\bf R}$ and ${\bf R}^{\prime}$, used
to describe the reaction are sketched in Fig. \ref{fig1}. We denote
the conjugate momenta by the corresponding ${\bf p}$ and use primes
for the final states. We neglect the Coulomb interaction and assume
that the target has zero spin.  With the assumptions of two
independent subsystems in the final state, we must find both elastic
and absorption halo-target differential cross sections as products
\cite{gar98} of participant-target (elastic or absorption) cross
sections and the averaged spectator overlap matrix element
\cite{gar96}, i.e.
\begin{eqnarray} 
& \frac{d^9\sigma _{el}^{(i)}({\bf P}^{\prime },{\bf 
p}_{jk}^{\prime },{\bf p}_{0i}^{\prime })}
{ d{\bf P}^{\prime } d{\bf p}_{jk}^{\prime } d{\bf p}_{0i}^{\prime } }
 =   \nonumber \\ 
& \frac{d^3\sigma _{el}^{(0i)}({\bf p}_{0i}
 \rightarrow  {\bf p}_{0i}^{\prime})} 
 {d{\bf p}_{0i}^{\prime }} \;
  \frac{1}{2 J+1} \sum_{M s_{jk}\Sigma_{jk}\Sigma_i } 
 |M_{s_{jk} \Sigma_{jk} \Sigma_i}^{JM}|^{2}  \label{eq5} \\
&  \frac{d^6\sigma _{abs}^{(i)}({\bf P}^{\prime },{\bf 
p}_{jk}^{\prime })}
{ d{\bf P}^{\prime } d{\bf p}_{jk}^{\prime } } = \nonumber \\ 
&   \sigma _{abs}^{(0i)}(p_{0i}) \;
  \frac{1}{2 J+1} \sum_{M s_{jk}\Sigma_{jk}\Sigma_i } 
 |M_{s_{jk} \Sigma_{jk} \Sigma_i}^{JM}|^{2}  \; ,
\label{eq6}
\end{eqnarray}
where $\frac{d^3\sigma _{el}^{(0i)}({\bf p}_{0i} \rightarrow {\bf
p}_{0i}^{\prime})} {d{\bf p}_{0i}^{\prime }}$ is the
participant-target differential elastic cross section, $\sigma
_{abs}^{(0i)}$ is the participant-target absorption cross section,
$\Sigma_i$ and $\Sigma^{\prime}_i$ are spin projections of halo
particle $i$ before and after the reaction, $s_{jk}$ and $\Sigma_{jk}
= \Sigma_{j} + \Sigma_{k}$ are total spin and projection of the halo
particles $j$ and $k$, $M_{s_{jk} \Sigma_{jk},\Sigma_i}^{JM}$ is the
overlap matrix element between initial and final states of the
spectator wave functions \cite{gar96} and $J$ and $M$ are total
angular momentum and projection on the beam direction of the initial
halo wave function.

In the experiment the core and a neutron are detected in coincidence
with velocities approximately equal to the beam velocity. Thus in
agreement with the participant-spectators approximation one neutron
reacts with the target without destroying or significantly affecting
the motion of the core and the other neutron.  The finite extension of
the projectile and the target therefore requires that only
configurations where the participant is sufficiently far away from the
spectators can contribute to the reaction.  We account for this by
omitting those geometric configurations in the initial wave function
where the participant ($n$) is closer to the two spectators ($n$ and
$c$) than cutoff distances $r_{nn}$ and $r_{nc}$ which are treated as
parameters. This shadowing effect substantially reduces the absolute
values of the cross sections \cite{gar98,hen97}.

{\it Observable and parameters.} Recently a new observable was
measured after fragmentation of a two-neutron halo nucleus on a carbon
target. The projectile interacts with the target and the angular
distribution of the relative momentum of the detected neutron-core
system is measured in a coordinate system with the $z$--axis along its
center of mass momentum \cite{chu97}. This process has a contribution
of about 70\% from absorption of the participant-neutron
\cite{coo93}. The remaining about 30\% arises from elastic scattering
of the participant-neutron which leaves two neutrons in the final
state. Subsequently the equal contributions ($\approx$ 15\%) from the
spectator-neutron (absorption like distribution) or the
participant-neutron are measured. The contribution from the
interaction of the core with the target is expected to be negligibly
small \cite{gar98}.

We shall first concentrate on the $\approx$ 85\% of the absorption
like distribution. Then the relative momentum is ${\bf p}_{jk}^\prime$
and the center of mass momentum is the conjugate final state momentum
${\bf p}_{i,jk}^\prime$ of the coordinate ${\bf r}_{i,jk}$, see
Fig. \ref{fig1}. The angular correlation is then computed from
eq.(\ref{eq6}) by integration over all momentum coordinates except the
angle $\theta$ between these ${\bf p}_{jk}^\prime$ and ${\bf
p}_{i,jk}^\prime$.

We consider the nuclei $^6$He (n+n+$\alpha$) and $^{11}$Li
(n+n+$^{9}$Li) with the wave function obtained by solving the Faddeev
equations in coordinate space \cite{gar96} with the potentials from
\cite{cob97}.  The resulting three-body wave functions have 88\% of
p$^2$- and 12\% of the neutron-core s$^2$-configurations for $^6$He
and 20\% of p$^2$- and 80\% of s$^2$ for $^{11}$Li. The binding
energies and the root mean square radii are (0.95 MeV, 2.45 fm) and
(0.305 MeV, 3.34 fm), respectively.  For the neutron-target
interactions we use non-relativistic optical potentials \cite{udi95}
with the phenomenological neutron--$^{12}$C parameterization EDAI-C12
\cite{coo93} valid for a range of neutron energies from 29 to 1040
MeV.  We include 35 partial waves in the calculations.

The binding energies and sizes of the initial states and the
experimental neutron and core momentum distributions are essentially
reproduced with these parameters \cite{gar96,gar97}. Furthermore,
shadowing parameters of 3 to 4 fm maintain or even improve the
agreement of these results for $^6$He and in addition the available
absolute two-neutron removal cross section is reproduced \cite{gar98}.

\begin{figure}
\psfig{file=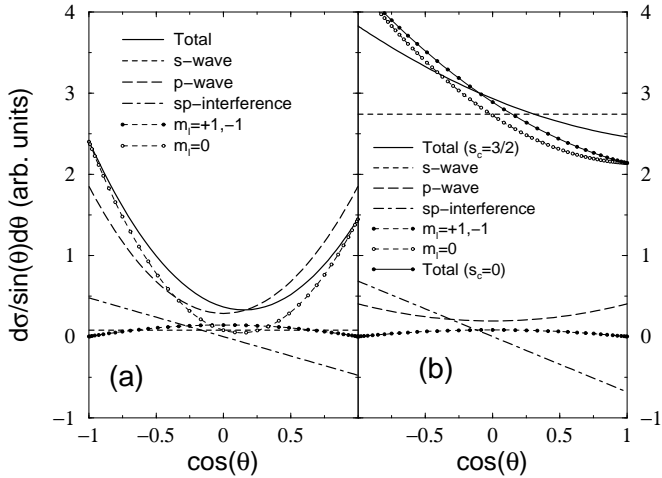,width=87mm,angle=-90,%
bbllx=35mm,bblly=18mm,bburx=200mm,bbury=245mm}
\vspace{3mm}
\caption{The absorption contribution, computed in the projectile
center of mass, from the spectator-neutron ($\approx$ 70\%) to the
neutron-core (spectators) angular correlation after fragmentation of
$^6$He (left) and $^{11}$Li (right) on $^{12}$C at 240 MeV/u.  The
shadowing parameters are equal $r_{nn}=r_{n\alpha}$ and 3 fm for
$^6$He and 4 fm for $^{11}$Li. We show the contributions from the
neutron-core (spectators) relative $s$ and $p$-waves, their
$sp$-interference and the three different projections on the $^5$He or
$^{10}$Li center of mass momentum which is very close to the beam
direction.  To obtain the contributions from the individual
projections we assumed zero core spin of $^{9}$Li ($s_c=0$) for
$^{11}$Li. The total result for $s_c=3/2$ is also shown (right).  The
curves are shown with the relative normalization obtained from the
initial wave function.}
\label{fig2}
\end{figure}

{\it Partial wave division.}  The initial three-body wave function
contains in our examples both $s^2$ and $p^2$ relative neutron-core
configurations. Removal of one neutron leaves the remaining
neutron-core system (spectators) in a mixture of $s$ and $p$-waves,
which after the absolute square and subsequent integration allow
diagonal $s$ and $p$-terms as well as an $sp$-interference term.  The
shapes of the resulting three angular distributions differ
substantially and the weighting, which of course is predicted by our
model, is decisive.  However, the experimentally preferred reaction
mechanism might turn out to be different.  To gain insight we show in
Fig. \ref{fig2} the contributions from these different partial waves.

The $s$-waves are angle-independent and insignificant for $^{6}$He and
dominating for $^{11}$Li. The $p$-waves vary symmetrically with angle
and the asymmetric $sp$-interference terms change sign for $\cos
\theta = 0$ and would therefore not contribute to the total cross
section obtained by integration over the angle $\theta$.  Thus the
angular correlation for $^{6}$He is essentially determined by the
contributions from the $p$-wave and it is therefore symmetric. For
$^{11}$Li the angular variation is essentially due to the
$sp$-interference and therefore very asymmetric.

We also in Fig. \ref{fig2} show the contributions arising from three
different angular momentum projections along the neutron-core
(spectators) center of mass momentum. For simplicity we assumed zero
$^{9}$Li-core spin ($s_c=0$) in these computations. The difference
between results for $s_c=0$ and $s_c=3/2$ is visible but not
substantial.  The asymmetry is for both nuclei again due to the
$sp$-interference appearing only for $m_l=0$, which contains part of
the $p$-wave and all the $s$-wave contributions. These terms with
$m_l=0$ exhibit a strong angular variation.  The terms with $m_l=\pm
1$, arising entirely from $p$-waves, are symmetric due to the lack of
$sp$-interference and vary relatively little with angle.

\begin{figure}
\psfig{file=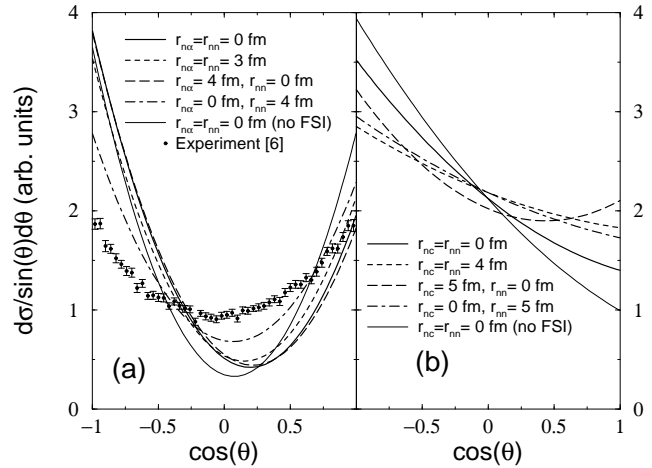,width=3.3in,angle=-90,%
bbllx=35mm,bblly=18mm,bburx=200mm,bbury=241mm}
\vspace{3mm}
\caption{The total absorption distribution for the same reactions as in
Fig. \ref{fig2}. The results for different shadowing parameters are
compared with the available measured distributions
\protect\cite{chu97}.  For comparison calculations without shadowing
and without final state interactions are shown. All calculated curves
are normalized to the same value.}
\label{fig3}
\end{figure}

{\it Effects of shadowing.} We compare in Fig. \ref{fig3} the measured
angular distribution with computations for various shadowing
parameters.  The first impression is that the computed and measured
distributions are quite different. Both the asymmetry and the
variation with angle are much larger for the computed curves.

Shadowing is simply removal of unwanted geometric configurations in
the initial wave function.  We first only remove the part of the
three-body wave function within a sphere around the core (long-dashed
curves), which implies predominant removal of the relative
neutron-core (spectators) $s$-states.  We obtain an essentially
unchanged distribution for $^{6}$He, since the $s$-wave only
contributes marginally ($\approx$ 12\%) and all the $p$-waves are
reduced by the same amount. The result is a reduction factor
independent of angle.  For $^{11}$Li this shadowing is more visible,
since the predominant removal of the dominating $s$-states now produce
a more $p$-like structure, i.e. more symmetric but still varying with
angle.

If we instead only remove a sphere around the spectator-neutron
(dot-dashed curves) we predominantly remove relative neutron-neutron
$s$-states, which must be transformed to the neutron-core Jacobi
coordinate system for the spectators.  For $^{6}$He the removed wave
function mostly consists of relative neutron-neutron $s$-waves with
one nodes (hyperspherical quantum number $K=2$). The transformed wave
function then has a smaller relative content of the dominating
$p$-waves and the $p$ and $sp$-contributions decrease relatively
compared to the $s$-wave contribution with a smaller asymmetry as the
consequence.  For $^{11}$Li both the removed and transformed wave
function is mostly $s$-waves without node ($K=0$). The reduction of
the dominating $s$-wave contribution then produces a more symmetric
$p$-like angular distribution.

The final state interaction is essential to reproduce the observed
narrow neutron momentum distribution \cite{gar96}. For the angular
correlation the effect is significant but still substantially smaller
than the difference to the measured distribution. The final state
interaction preserves the total angular integrated contribution from
each partial wave.  The contribution from the neutron-core
(spectators) relative $s$-wave is angle independent. However,
including the final state interaction and combining with the $p$-wave
contributions the asymmetry for $^{6}$He or $^{11}$Li increase or
decrease, respectively.

{\it Dependence on spectator excitation energy.}  We have assumed that
the reaction is dominated by one-neutron absorption while the
remaining neutron-core (spectators) system is undisturbed. The
argument is that this high-energy process must be very fast compared
to the time scale of the intrinsic motion of the halo particles. The
reaction has occurred before the remaining halo particles can change
their relative motion and for example select an excited state which in
the present cases must be two-body resonances or virtual states.  This
is the sudden approximation used so successfully for many other
observables \cite{gar96,gar97,cob97}.  In this picture all continuum
states are populated with the probability of occurrence in the initial
wave function. However, it is conceivable that a different continuum
population arises in these reactions and in particular when the beam
energy is substantially reduced below the values in the present
experiments.

\begin{figure}
\psfig{file=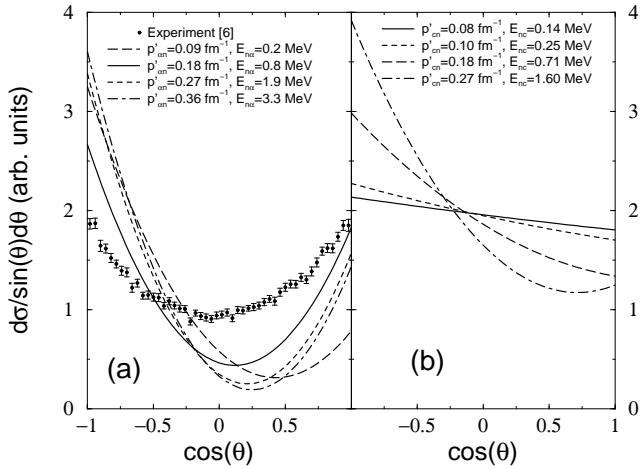,width=3.3in,angle=-90,%
bbllx=35mm,bblly=18mm,bburx=200mm,bbury=241mm}
\vspace{3mm}
\caption{The total absorption distribution for the parameters and
reactions in Fig. \ref{fig2}. The results are given for various
relative neutron-core (spectators) excitation energies $E_{n\alpha}$
and $E_{nc}=p_{nc}^{\prime2}/2\mu$.  All calculated curves are
normalized to the same value. }
\label{fig4}
\end{figure}

To investigate the sensitivity of the angular correlation to the
continuum population we show in Fig. \ref{fig4} the results when only
a state with a given relative neutron-core (spectators) excitation
energy is populated in the final state.  This energy is varied from
zero and across the energies of the lowest resonances \cite{cob97}.
We find for $^{6}$He that all the distributions vary too much with
angle. The lowest energy of 0.2 MeV exhibits a large asymmetry
indicating a relatively large admixture of $s$-waves still far from
the measured distribution. The higher energies all produce similar
distributions even when the energy matches the $p$-resonance at 0.77
MeV. The continuum two-body states are essentially $p$-waves due to
the low-lying resonance and the distributions in fact resemble the
$p$-wave distribution from Fig. \ref{fig2}.  Thus, no weighted average
of these continuum states can reproduce the measured distribution.

For $^{11}$Li the low-lying virtual $s$-states influence the angular
distribution. We find a rather flat and asymmetric distribution at low
energy, where the $s$-waves dominate completely. As the excitation
energy increases the $p$-waves contribute more and more. This produces
an increased asymmetry due to the interference term and at higher
energies the $p$-wave contribution is clearly pulling the distribution
towards symmetry.

\begin{figure}
\psfig{file=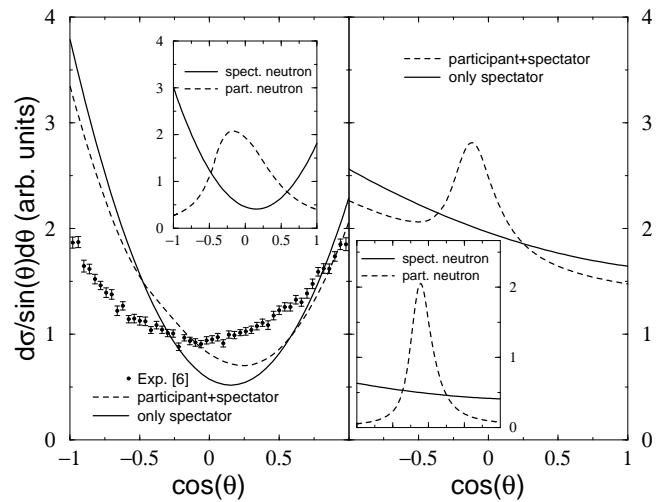,width=87mm,angle=-90,%
bbllx=35mm,bblly=22mm,bburx=200mm,bbury=241mm}
\vspace{3mm}
\caption{The angular distribution for the parameters and reactions in
Fig. \ref{fig3}. The insets show the two contributions from the
scattering process with two neutrons and a core in the final
state. Each of these contributions amounts to about 15\%, i.e. when
the participant (dashed) and the spectator-neutron (solid),
respectively are measured. The solid curves in the external parts give
the total contributions when the approximately 15\% from the
participant-neutron is completely neglected, i.e. we show the sum of
the solid curve ($\approx$ 15\%) from the insets and the total
contribution ($\approx$ 70\%) from Fig. \ref{fig2}. The dashed curves
in the external parts show the sum of the solid curves ($\approx$
85\%) and the dashed curves ($\approx$ 15\%) from the inset.  }
\label{fig5}
\end{figure}

{\it Contribution from the participant-neutron.}  If the
participant-neutron is scattered by the target instead of being
absorbed it may still be detected and contributes then about 15\% to
the measured cross section. The estimate assumes that this neutron
arrives within the forward angle where the detection takes place. This
relatively small contribution is for technical reasons only estimated
approximately. We first approximate the motion of the center of mass
of the neutron-core (participant-spectator) after the reaction to be
in the direction of the beam.  Then the participant-neutron and
spectator-core relative momentum is approximated, as for an infinitely
heavy core, by the momentum of this participant-neutron relative to
the projectile center of mass.  The heavier the core and the higher
the beam energy the better the approximations. Finally we obtain the
contribution from eq.(\ref{eq5}) by integration over all momenta
except the angle between these two momenta. The result is shown
Fig. \ref{fig5}.

The contribution from the participant-neutron is almost symmetric and
peaked at $\cos \theta = 0$, see the inset of Fig. \ref{fig5}. This
behavior is in striking contrast to the contribution from the
spectator-neutron which has a minimum at $\cos \theta = 0$ for
$^{6}$He and is rather flat for $^{11}$Li.  The peak corresponds to a
preferred direction perpendicular to the beam direction as obtained
for forward scattering of a neutron on a target expressed in the rest
system of the neutron.  Such a distribution is broadened by the
intrinsic motion of the scattered participant-neutron with respect
to the center of mass of the two spectator-particles.  Thus, the
higher the beam energy and the smaller the halo binding energy, the
narrower the distribution.

For $^{6}$He the contribution from the participant-neutron only
changed the total distribution very little, see Fig. \ref{fig5}.  For
$^{11}$Li a peak appears due to the rapid angular variation of the
contribution from the scattered participant-neutron compared to the
almost constant background from the dominating $s$-waves, see
Fig. \ref{fig5}.  Before a direct comparison with measurements it is
worth keeping in mind that this result is obtained first as a crude
estimate and furthermore with all participant-neutron contributions
included independent of possible additional experimental
selection.

{\it Conclusion.}  We computed the recently measured angular
correlation for breakup reactions of halo nuclei $^{6}$He and
$^{11}$Li, i.e. the halo nucleus interacts at high energy with the
target and the angular distribution of the relative momentum of the
detected $^{5}$He and $^{10}$Li neutron-core system is measured in a
coordinate system with the $z$--axis along the center of mass
momentum. We used a model which successfully describes essentially
all other three-body observables for such systems.

We investigated the contributions from different partial waves of the
relative motion of the detected neutron-core system and found
constant, symmetric and asymmetric distributions arising respectively
from diagonal $s$, diagonal $p$ and $sp$-interference terms.  The zero
angular momentum projection on the direction of the $^{5}$He and
$^{10}$Li momenta produces by far the largest variation with angle.

Exciting continuum neutron-core states of a given energy in the
reaction with the subsequent decay of these states might be a possible
reaction mechanism which however is not supported by the present
analysis. The energy is too high to allow time to select specifically
for example the two-body resonance states and agreement with the
available data is furthermore not improved.

Two processes are possible, i.e. absorption or scattering of one
neutron (participant) while the other neutron and the core
(spectators) continue undisturbed. The detected neutron-core system
may consist of either participant or spectator-neutron and the
resulting angular distributions are qualitatively different. The
process where the participant-neutron is detected contributes by less
than the 15\% predicted by the optical model, since only a narrow cone
around the beam direction is selected in the experiment.

The correlations for $^{6}$He and $^{11}$Li are qualitatively
different and the computed distribution vary too strongly compared to
the data for $^{6}$He. Provided the experimental correlation is
correct this indicates that a subtle reaction mechanism is at
work. From the present investigation this can be either a missing
constant background for example due to suppression of $p$-waves or a
preferred selection of $s$-waves or a relative suppression of zero
angular momentum projections.

{\bf Acknowledgments.} We thank K. Riisager for continuous discussions
and suggestions and T. Aumann and L. Chulkov for a number of
clarifying remarks in connection with the experimental data.


\begin{thebibliography}{99}

\bibitem{han95} P.G. Hansen, A.S. Jensen and B. Jonson, 
 Ann. Rev. Nucl. Part. Sci. {\bf 45}, 591 (1995)

\bibitem{zhu93}  M.V. Zhukov, B.V. Danilin, 
D.V. Fedorov, J.M. Bang, I.J. Thompson and J.S. Vaagen, 
Phys. Reports {\bf 231}, 151 (1993).

\bibitem{gar96} E. Garrido, D.V. Fedorov and A.S. Jensen, Phys. Rev.
{\bf C55}, 1327 (1997).

\bibitem{gar97} E. Garrido, D.V. Fedorov and A.S. Jensen, 
Nucl. Phys.  {\bf A617}, 153 (1997).

\bibitem{gar98} E. Garrido, D.V. Fedorov and A.S. Jensen, 
{\it Preprint nucl-th/9803056}; Europhysics Lett., in press. 

\bibitem{chu97} L.V. Chulkov et al., Phys. Rev. Lett. {\bf 79}, 201 (1997).

\bibitem{ban67} J. Bang and C.A. Pearson, 
Nucl. Phys.  {\bf A100}, 1 (1967).

\bibitem{udi95} J.M. Ud\'{\i}as, P. Sarriguren, E. Moya de Guerra,
E. Garrido and J.A. Caballero,   Phys. Rev.  {\bf C51}, 3246 (1995).

\bibitem{hen97} G.F. Bertsch, K. Hencken and H. Esbensen, 
 Phys. Rev.  {\bf C57}, 1366 (1998).

\bibitem{coo93} E.D. Cooper, S. Hama, B.C. Clark and R.L. Mercer,
  Phys. Rev. {\bf C47}, 297 (1993).

\bibitem{cob97} A. Cobis, D.V. Fedorov and A.S. Jensen, 
Phys. Rev. Lett. {\bf 79}, 2411 (1997); Phys. Lett.  {\bf B424}, 1 (1998).

\end{thebibliography}
\end{document}